\documentclass[aps,prc,reprint,groupedaddress]{revtex4-2}
\newcommand*{\rom}[1]{\expandafter\@slowromancap\romannumeral #1@}
\makeatother
\usepackage{amsmath}
\usepackage{graphicx}
\usepackage{dcolumn}
\usepackage{bm}
\usepackage{algpseudocode}
\usepackage[caption=false]{subfig}
\usepackage{url}
\usepackage{cases}
\usepackage{longtable}
\usepackage{float}
\usepackage{booktabs}

\begin{document}


\title{Simultaneous improvements of nuclear mass and charge radius predictions using multi-task Gaussian process approaches}


\author{Weihu Ye}
\affiliation{School of Physics and Optoelectronics, South China University of Technology, Guangzhou 510641, China}
\author{Niu Wan}
\email{wanniu@scut.edu.cn}
\affiliation{School of Physics and Optoelectronics, South China University of Technology, Guangzhou 510641, China}
\newcommand{\RNum}[1]{\uppercase\expandafter{\romannumeral #1\relax}}


\date{\today}

\begin{abstract}
A multi-task Gaussian process (GP) machine learning model is introduced to simultaneously predict two important nuclear observables across the nuclear chart, namely nuclear masses and charge radii. Utilizing 12 physical input features, our multi-task GP consistently outperforms single-task learning, achieving overall root-mean-square deviations of 0.136 MeV for masses and 0.007 fm for charge radii. The good performance of the present model is confirmed by three complementary validations, namely various fractions for training and testing data, further extrapolations for newly reported nuclei far from stability, and popular Garvey–Kelson mass relations. The correlations between the two observables are explicitly analyzed within the multi-task learning framework. Furthermore, by employing the SHapley Additive exPlanations (SHAP) method, we interpret the importance of different features for mass and radius predictions across distinct nuclear regions. These results demonstrate the effectiveness of the multi-task GP approach for high-accuracy nuclear property predictions.
\end{abstract}

\maketitle


\section{INTRODUCTION}
Nuclear mass and charge radius are fundamental properties encoding essential nuclear structure information, such as shell closures \cite{sahin2015shell,Lunney}, deformation \cite{moller2016nuclear}, and symmetry energy \cite{huang2023correlation}. Meanwhile, these two observables also impose crucial constraints on numerous astrophysical processes, such as the rapid neutron capture process and the composition of the neutron star crust \cite{arnould2007r,mumpower2016impact}. Recent experimental progress has significantly expanded the datasets of nuclear masses and charge radii, especially for nuclei around the $\beta$-stability line. The datasets for nuclear mass and charge radius have been systematically compiled in the latest Atomic Mass Evaluation (AME2020) \cite{wang2021ame} and the CR2013 table \cite{angeli2013table}, respectively, with more than 2400 known nuclear masses and 900 measured charge radii. However, there are still a large number of nuclear masses and charge radii without experimental measurements, especially for nuclei far from stability. Therefore, reliable theoretical models are very necessary to predict these quantities with high accuracy across the nuclear chart.

Most macroscopic nuclear mass models and macroscopic-microscopic approaches, such as liquid-drop model \cite{bethe1936nuclear,kirson2008mutual}, finite-range droplet model \cite{moller2012new}, Weizs\"acker–Skyrme-type models \cite{wang2010modification,wang2014surface}, and Duflo–Zuker mass models \cite{duflo1995microscopic}, have been developed primarily to reproduce the nuclear masses. These models achieve root-mean-square (rms) deviations from 0.3 MeV to 5 MeV with respect to the experimental mass data \cite{MinWang}. However, the nuclear charge radius is usually not directly involved in these models. In contrast, microscopic models based on nuclear energy density functionals, such as Skyrme Hartree-Fock-Bogoliubov \cite{bender2003self,goriely2013further} and relativistic mean-field models \cite{meng2006relativistic,vretenar2005relativistic,liang2013feasibility}, can simultaneously describe nuclear masses and charge radii within a unified framework. However, their predictions of nuclear masses usually exhibit the rms deviations larger than 1 MeV \cite{agbemava2014global,Afanasjev,erler2012limits}, while the theoretical charge radii yield rms deviations about 0.02–0.05 fm \cite{afanasjev2021global,goriely2007further}. Hence, simultaneous description with high accuracy for both nuclear mass and charge radius is still required within a unified theoretical framework.

One potential strategy to satisfy this description is the application of multi-task machine learning (ML) approach that explicitly utilizes the correlations between nuclear masses and charge radii. In recent years, a variety of ML techniques have been introduced to nuclear physics \cite{wang2023machine,niu2019comparative,he2023machine,boehnlein2022colloquium,carleo2019machine,gao2021machine,karagiorgi2022machine,wu2022nuclear,wu2024nuclear,niu2022nuclear,Piekarewicz,sharma2022learning,li2022deep,huang2025validation,li2025investigation}, such as Gaussian process (GP) models \cite{ye2025understanding,yuksel2024nuclear}, Mixture density networks \cite{lovell2022nuclear,021301,li2024atomic}, and Bayesian neural networks \cite{Utama,Piekarewicz,Niu}, which have been successfully applied to studies of nuclear masses or charge radii, resulting in improved accuracy over traditional theoretical models. Recent progress reveals that the rms deviation for nuclear masses achieved by most ML approaches can reach 0.2 MeV \cite{lovell2022nuclear,021301,li2024atomic,Utama,Piekarewicz,Niu}, and the rms deviation for charge radii can be as small as around 0.01 fm \cite{cao2023predictions,dong2023nuclear,su2023progress}. However, most of these ML techniques focus on predicting a single task, either the nuclear mass or the charge radius, without utilizing the correlations between them. Recently, several studies have applied multi-task learning frameworks to nuclear data, such as the simultaneous modeling of nuclear masses and separation energies \cite{ming2022nuclear,wu2022multi}, demonstrating the potential benefits of joint learning. However, multi-task ML approaches for simultaneously and systematically predicting both nuclear masses and charge radii are rarely involved. One may easily confuse multi-task learning with multi-objective learning \cite{021301}. The latter has proven effective in improving the accuracy of models for a single observable, such as nuclear masses, by optimizing several loss functions simultaneously, but it does not capture correlations across different observables. In contrast, multi-task learning models jointly learn multiple outputs and explicitly leverage their interdependence, which is essential for consistent predictions of nuclear masses and charge radii.

In this work, we introduce a multi-task GP framework that enables the simultaneous modeling of both nuclear masses and charge radii. By utilizing the correlations between these two tasks, our approach provides a simultaneous description and improves the predictive accuracy for both tasks compared to each single-task learning. Furthermore, we perform three complementary generalization tests to assess the reliability of present model, including its stability with random train–test partitions, further extrapolation performance for newly reported nuclei, and good consistency with the well-known Garvey–Kelson (GK) relations \cite{garvey1969set,Barea}. The structure of this paper is as follows. In Section II, we present the theoretical description of GP and its application to multi-task learning. Feature selection, data standardization, and model training are also introduced here. In Section III, the single-task and multi-task GP models with different feature sets are compared and their predictive performances for nuclear masses and charge radii are estimated by three complementary tests of generalization. Moreover, the correlations between these two tasks are also analyzed via covariance analysis, and the importance of different features for nuclear masses and charge radii is also interpreted across different nuclear regions. Finally, the main findings and conclusions are summarized in the last section.

\section{THEORETICAL FRAMEWORK AND METHODS}
\subsection{Multi-task Gaussian process}
\label{MTGP}
The Gaussian process (GP) is a nonparametric ML method \cite{2006,schulz2018tutorial,wang2023intuitive}, which has been widely used for modeling nuclear observables \cite{yuksel2024nuclear,Neuf2020}. The GP can flexibly fit data and incorporate prior knowledge via the covariance (kernel) function, allowing the model to capture complex correlations among nuclear observables. A commonly adopted kernel in GP models is the Mat\'{e}rn-3/2 kernel \cite{2006,schulz2018tutorial,wang2023intuitive}, which captures smooth correlations in the data. The expressions of the Mat\'{e}rn-3/2 can be written as:
\begin{equation}
 k(x_i, x_j) = \left(1 + \frac{\sqrt{3}\,\lVert x_i - x_j \rVert}{\ell}\right)
    \exp\left(-\frac{\sqrt{3}\,\lVert x_i - x_j \rVert}{\ell}\right),
\end{equation}
where \(\lVert \mathbf{x}_i - \mathbf{x}_j\rVert\) denotes the Euclidean distance between feature vectors \(\mathbf{x}_i\) and \(\mathbf{x}_j\), and $\ell$ is the characteristic length scale controlling the smoothness of the function. To enhance numerical stability, a standard white-noise kernel is included to account for random fluctuations in the data \cite{yuksel2024nuclear,ye2025understanding}. Within this noise range, the GP model automatically converges and optimizes the noise contribution during training. For the involved tasks in this work, $x$ is taken as a vector of 12 physical quantities (discussed in the next subsection) relevant to the nuclear structure. In this case, the corresponding Mat\'{e}rn-3/2 kernel can be further represented as
\begin{equation}
\begin{aligned}
k(x_i, x_j) &= \prod_{k=1}^{12} 
\left(1 + \frac{\sqrt{3}\,\lvert x_{i,k} - x_{j,k} \rvert}{\ell_k}\right) \\
&\quad \times \exp\!\left(-\frac{\sqrt{3}\,\lvert x_{i,k} - x_{j,k} \rvert}{\ell_k}\right),
\end{aligned}
\label{eq10}
\end{equation}
where $x_{i,k}$ and $x_{j,k}$ denote the $k$-th physical feature of nuclei $i$ and $j$, respectively. 

The standard GP method is usually applied for a single task. However, multiple-correlated tasks have the potential to simultaneously improve predictive accuracies, such as nuclear masses and charge radii involved in the present work. In order to achieve the multi-task GP, we employ the widely used framework of the intrinsic coregionalization model (ICM) \cite{alvarez2012kernels,bonilla2007multi}, which enables simultaneous learning of several related quantities with a shared input feature set. To consider multiple tasks simultaneously, the ICM approach introduces a covariance function ($K_{\text{ICM}}$) that captures both the similarity among the input features and the correlation among different tasks. The explicit formulation of $K_{\text{ICM}}$ is given below.

Suppose that there are $n_m$ measured nuclear masses and $n_r$ measured charge radii, giving data points with a total number $n = n_m + n_r$ and two tasks with $q = 2$. The full covariance matrix $K_{\text{ICM}}$ combines the input kernel matrix $K_X$ and the task covariance matrix $B$, where the input kernel matrix $K_X$ ($n \times n$) describes the similarity between nuclei based on the Mat\'{e}rn-3/2 kernel defined in Eq.~(\ref{eq10}) with
\begin{equation}
K_X =
\begin{bmatrix}
k(x_1, x_1) & k(x_1, x_2) & \cdots & k(x_1, x_n) \\
k(x_2, x_1) & k(x_2, x_2) & \cdots & k(x_2, x_n) \\
\vdots & \vdots & \ddots & \vdots \\
k(x_n, x_1) & k(x_n, x_2) & \cdots & k(x_n, x_n)
\end{bmatrix},
\end{equation}
and the coregionalization matrix $B$ ($q \times q$) simulates the correlation across different tasks with
\begin{equation}
B =
\begin{bmatrix}
  B_{11} & B_{12} \\
  B_{21} & B_{22}
\end{bmatrix}.
\label{eq2}
\end{equation}
Here, $B_{11}$ and $B_{22}$ are the prior variances for the nuclear mass and charge radius tasks, respectively, which describe the overall variability of each quantity across different nuclei.  The term $B_{21}=B_{12}$ represents the covariance between the two tasks. All the elements in the $B$ matrix are treated as free hyperparameters during training. 

Each element of the full covariance matrix $K_{\text{ICM}}$ for all tasks and data points is computed by combining the input kernel matrix $K_X$ with the task covariance matrix $B$. Explicitly, the matrix element $K_{\text{ICM}(i,t_i),(j,t_j)}$ connecting the $i$-th data point from task $t_i$ with the $j$-th data point from task $t_j$ is given by
\begin{equation}
K_{\text{ICM}(i,t_i),(j,t_j)} = k(x_i, x_j) B_{t_i t_j},
\label{eq11}
\end{equation}
where the task indices $t_i, t_j \in \{1,2\}$ denote specific tasks with 1 for nuclear mass and 2 for charge radius. By partitioning the data points according to their respective tasks, the full covariance matrix $K_{\text{ICM}}$ can be explicitly written in block form as:
\begin{equation}
K_{\text{ICM}} =
\begin{bmatrix}
B_{11}\,K_X^{(m,m)} & B_{12}\,K_X^{(m,r)}\\[6pt]
B_{21}\,K_X^{(r,m)} & B_{22}\,K_X^{(r,r)}
\end{bmatrix},
\end{equation}
where $K_X^{(m,m)}$ ($n_m \times n_m$) contains kernel evaluations between only nuclear masses, $K_X^{(r,r)}$ ($n_r \times n_r$) contains kernel evaluations between only charge radii, and $K_X^{(m,r)} = [K_X^{(r,m)}]^\top$ ($n_m \times n_r$) contains kernel evaluations between nuclear mass and charge radius. Thus, the diagonal blocks encode covariances within each task, while the off-diagonal blocks represent covariances between different tasks. This structure enables the model to capture both input similarities and inter-task relationships, allowing effective information sharing across tasks.

To quantitatively characterize the degree of correlation between the two tasks (nuclear mass and charge radius), the Pearson correlation coefficient $r$ associated with the task covariance matrix $B$ is usually considered, which can provide a direct estimation of the strength and direction for the linear relationship between the two tasks \cite{2006}:
\begin{equation}
r = \frac{B_{12}}{\sqrt{B_{11} B_{22}}}.
\label{eq3}
\end{equation}

\subsection{Feature space and numerical details}
\label{features}
In our recent work, we systematically checked the effects on the predictions of theoretical masses with 5, 9, and 12 input features within a single-task ML model \cite{ye2025understanding}, where the last set can achieve the best accuracy. In this work, the same latter two sets M9 and M12 are used to further check the effects on the performance of the multi-task ML model. The features are summarized in Table~\ref{tab:1}, where $N$, $Z$, and $A$ denote the neutron number, proton number, and mass number, respectively. The terms $A^{2/3}$ and $A^{1/3}$ represent the surface term in mass models and a quantity related to the charge radius, respectively. The feature $(N-Z)/A$ is the nuclear isospin asymmetry term, while $Z(Z-1)/A^{1/3}$ corresponds to the Coulomb term. $N_{eo}$ and $Z_{eo}$ describe the even-odd effects for neutrons and protons with values of 0 for even and 1 for odd numbers. $v_n$ and $v_p$ are the valence neutron and proton numbers defined as the distance from the nearest magic number ($N, Z = 8, 20, 28, 50, 82, 126, 184$). The final feature $P = v_n v_p/(v_n + v_p)$ is the Casten factor \cite{casten1996evolution}, which quantifies the coupling between valence nucleons.
\begin{table}[htbp]
\caption{Two sets of features for the ML model.}
\label{tab:1}
\centering
\begin{tabular}{cl}
\toprule
Feature set & Feature list\\
\midrule
M9  & $N$, $Z$, $A$, $A^{2/3}$, $A^{1/3}$, $(N-Z)/A$, $Z(Z-1)/A^{1/3}$,\\
    & $N_{eo}$, $Z_{eo}$ \\
M12 & $N$, $Z$, $A$, $A^{2/3}$, $A^{1/3}$, $(N-Z)/A$, $Z(Z-1)/A^{1/3}$,\\
    & $N_{eo}$, $Z_{eo}$, $v_{n}$, $v_{p}$, $P$  \\
\bottomrule
\end{tabular}
\end{table}

In the present work, we use 2340 experimental nuclear masses from the AME2020 \cite{wang2021ame} and 880 experimental nuclear charge radii from the CR2013 \cite{angeli2013table} with $Z, N \geq 8$. For all calculations, 70\% of the data are randomly assigned to the training set, while the remaining 30\% are for the testing set. The rms deviation (denoted by $\sigma$) is used to quantify the accuracy between the predicted values and the experimental data for both nuclear masses and charge radii. It is defined as
\begin{align}
\sigma = \sqrt{ \frac{1}{n} \sum_{i=1}^{n} \left( y_{i}^{\mathrm{pred}} - y_{i}^{\mathrm{exp}} \right)^2 },
\label{EQ:1}
\end{align}
where $y_{i}^{\mathrm{pred}}$ and $y_{i}^{\mathrm{exp}}$ represent the predicted and experimental values (for nuclear mass or charge radius), and $n$ is the number of the involved nuclei.

\subsection{Standardization of nuclear mass and charge radius data}
It should be noted that nuclear masses and charge radii are measured in different physical units and have different orders of magnitude. To make the two quantities comparable, each variable is standardized by using the Z-score normalization \cite{hastie2009elements}, which is defined as
\begin{align}
    y^{(\mathrm{s})}_i &= \frac{y_i - \mu}{\tau} \quad\text{with} \\
    \mu &= \frac{1}{n}\sum_{i=1}^{n} y_i \quad\text{and} \notag \quad \tau = \sqrt{\frac{1}{n-1}\sum_{i=1}^{n}(y_i-\mu)^2}, \notag
\end{align}
where $y^{(\mathrm{s})}_i$ denotes the standardized value for the nuclear mass or charge radius, $n$ is the number of data points, $\mu$ is the mean value, and $\tau$ is the standard deviation. In this way, both nuclear mass and charge radius are transformed into dimensionless quantities on the same scale. After model training, the standardization is reversed to recover the physical values of nuclear mass and charge radius.

\section{RESULTS AND DISCUSSION}
\label{sec:1}

\subsection{Performance comparison between single-task learning and multi-task learning models}
\label{performance}
\begin{figure*}
    \centering
    \includegraphics[width=18cm]{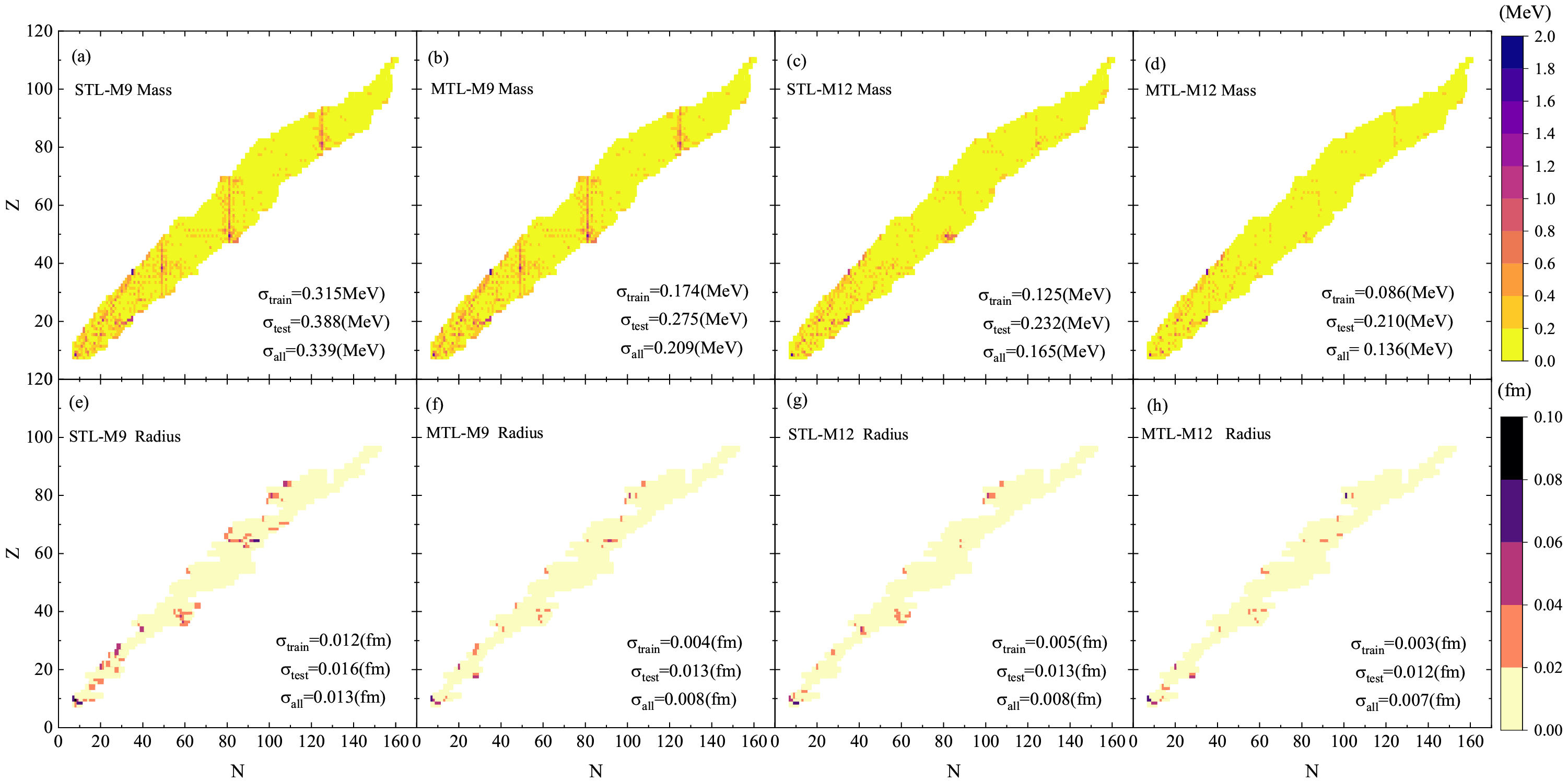}
    \caption{\label{fig1} Absolute errors between model predictions and experiment for nuclear masses and charge radii. Upper panels (a, b, c, d) show mass prediction results for the STL and MTL models, while lower panels (e, f, g, h) show charge radius prediction results for the STL and MTL models. Rms deviations calculated by Eq.~(\ref{EQ:1}) for the training, testing, and total sets are listed in each panel. Color bars indicate the magnitude of the absolute error (in MeV or fm). See text for details. }
\end{figure*}
We first evaluate the performance of single-task learning (STL) and multi-task learning (MTL) GP models using both M9 and M12 input feature sets. Fig.~\ref{fig1} shows the absolute errors between ML predictions and experimental data for nuclear masses and charge radii as well as the rms deviations calculated by Eq.~(\ref{EQ:1}) for the training ($\sigma_{train}$), testing ($\sigma_{test}$), and full ($\sigma_{all}$) datasets. In panel (a), the single-task learning model using 9 features (STL-M9) results in rms deviations of $\sigma_{\rm train}=0.315$~MeV and $\sigma_{\rm test}=0.388$~MeV, indicating not only the achievement of good accuracy on the training set but also the excellent generalization capability. The overall accuracy $\sigma_{\rm all}=0.339$~MeV is comparable to that of popular macroscopic-microscopic nuclear mass models. In contrast, as shown in panel (b), by simultaneously considering nuclear masses and charge radii, the multi-task learning model with 9 input features (MTL-M9) achieves a lower overall rms deviation of $\sigma_{\rm all}=0.209$~MeV with $\sigma_{\rm train}=0.174$~MeV and $\sigma_{\rm test}=0.275$~MeV for the training and testing sets, respectively. This corresponds to an improvement in accuracy of approximately 38\% compared to the STL-M9 result. 

In addition, as can be seen from panels (a) and (b), the largest mass deviations are generally located with nuclei near the shell closures. As shown in panels (c) and (d), with the inclusion of the last three shell-related features into the M12 set, these systematic deviations are further reduced, with $\sigma_{\rm all}=0.165$~MeV for STL-M12 and $\sigma_{\rm all}=0.136$~MeV for MTL-M12, respectively. This suggests that the additional features can effectively capture the shell effects in nuclear masses. Furthermore, the MTL-M12 also achieves a lower overall deviation by about 17\% compared to the STL-M12. The comparison between STL and MTL models with both the M9 and M12 feature sets demonstrates that incorporating charge radius information through MTL can significantly enhance the accuracy of mass predictions. In recent years, several advanced machine-learning mass models have reached accuracies close to or even below 100 keV \cite{niu2018high,niu2022nuclear,021301,wu2021nuclear}, and our MTL model achieves a comparable level of precision.

For charge radii predictions, the STL-M9 model achieves rms deviations of $\sigma_{\rm train}=0.012$~fm, $\sigma_{\rm test}=0.016$~fm, and $\sigma_{\rm all}=0.013$~fm. The MTL-M9 model reduces these to $\sigma_{\rm train}=0.004$~fm, $\sigma_{\rm test}=0.013$~fm, and $\sigma_{\rm all}=0.008$~fm, representing a 38\% reduction in the overall deviation. A similar trend is observed for the M12 feature set: STL-M12 produces $\sigma_{\rm all}=0.008$~fm, while MTL-M12 also obtains a lower value of $\sigma_{\rm all}=0.007$~fm. Notably, the accuracy of $0.007$~fm achieved by MTL-M12 compares favorably with that of recent ML models. Although the testing error $\sigma_{\rm test}=0.012$~fm is slightly larger than the training error, it remains fully consistent with the accuracy of recent ML models, such as Bayesian neural networks \cite{dong2023nuclear,dong2022novel}, kernel ridge regression \cite{ma2022improved}, and convolutional neural networks \cite{su2023progress}. From an experimental standpoint, this accuracy is comparable to the uncertainty of charge-radius measurements (typically $\pm 0.005$--$0.010$~fm) \cite{angeli2013table}. This indicates that the model’s predictive performance is close to the experimental limit and thus represents a realistic and reliable generalization. Overall, these results suggest that the present MTL framework can capture the correlation between nuclear mass and charge radius, resulting in mutual improvements in the prediction accuracy of both properties.

\subsection{Generalization capability of the models}
\label{validation}
A persistent challenge in applying machine learning to nuclear physics is its performance outside the training domain. Although ML models can reproduce known experimental observables with high accuracy, a critical test of reliability is whether they deliver consistent predictions for nuclei not yet measured experimentally or located far from the valley of stability. In this subsection, we assess the generalization capability of the present ML models using three complementary tests. The first examines the dependence of model accuracy on the fraction of training data. The second compares predictions with newly available experimental data on nuclear masses and charge radii that were not included in training. The third evaluates the consistency of predicted masses via the GK relations, which impose stringent local constraints among neighboring nuclei. Together, these tests provide a comprehensive appraisal of the robustness of the ML framework in both global extrapolation and local structural correlations.

\begin{figure}
 \includegraphics[width=8cm]{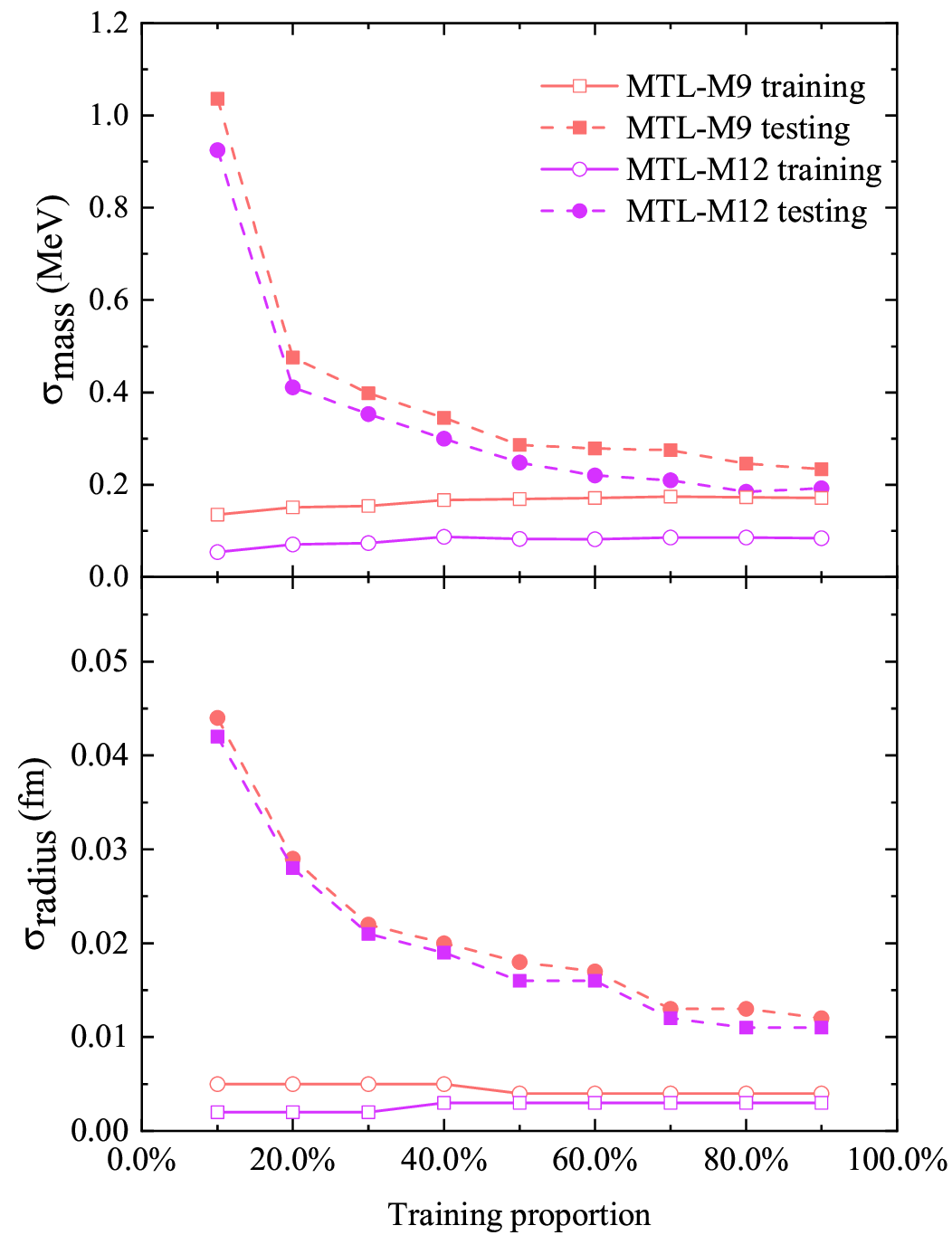}
\caption{\label{figover} Rms deviations of masses (top) and charge radii (bottom) with respect to the AME2020 and CR2013 datasets as functions of the training proportion, for the MTL-M9 and MTL-M12 models.}
\end{figure}

\subsubsection{Generalization with different training fractions}
It is well known that the model performance depends on the amount of training data. Therefore, varying the training fraction is a simple way to test the model stability and the data efficiency. Fig.~\ref{figover} shows the variation of the accuracy on the fraction of nuclei used for training from 10\% to 90\%. Consequently, the fraction for the testing set is reversed from 90\% to 10\%. 

It can be seen from Fig.~\ref{figover} that for nuclear mass, the training rms values are always small and nearly constant across different training fractions. This is because the model is always sufficient to learn the available data, indicating the good capacity of the model. In contrast, the testing rms decreases steadily as more nuclei are included in the training set, reflecting the decreasing overfitting and increasing generalization with larger training samples. The most significant reduction occurs when the training fraction increases from 10\% to about 50\%, while further improvements become minor once the fraction exceeds roughly 60\%. This saturation suggests that the main systematic trends in the data have already been well captured and the models achieve stable predictive accuracy beyond this point. In addition, MTL-M12 consistently outperforms MTL-M9 for both training and testing cases, indicating the advantage of including the shell-related features in the M12 set. For charge radius, the dependence on the training fraction follows a similar tendency that the training rms values are always stable while the testing rms values decrease with more training data and become stable at higher fractions. This suggests that both MTL-M9 and MTL-M12 achieve good stability for both nuclear masses and charge radii.

\subsubsection{Generalization with newly measured data}
Besides the training and testing sets constructed from AME2020 and CR2013 in Sec.~\ref{features}, we further introduce an extrapolation set to quantify the predictive power of the present ML models. The extrapolation set comprises newly reported experimental values after the two systematic evaluation tables. Specifically, we collected 21 newly measured nuclear masses \cite{silwal2022summit,giraud2022mass,izzo2021mass,mukul2021examining,paul2021mass,orford2022searching,porter2022investigating,xing2023isochronous,jaries2023high,xian2024mass,wang2024mass,jaries2024probing,kimura2024comprehensive,ireland2025high,mukai2025evidence,leistenschneider2021precision,beck2021mass,li2022first,ge2024high,mougeot2021mass} and 153 charge radii \cite{Li2021,Cubiss2023,MalbrunotEttenauer2022,Geldhof2022,Barzakh2021,Koenig2024,DayGoodacre2021}. Table~\ref{tab:newdata} summarizes the rms deviations for the two multi-task learning models, namely MTL-M9 and MTL-M12. 

As shown in Table~\ref{tab:newdata}, the rms deviations of the nuclear masses for the extrapolation set are consistent with the results of the testing set in Fig.~\ref{fig1}. In particular, MTL-M12 gives $\sigma_{\text{mass}}$=0.221 MeV for the new mass data, very similar to the value 0.210 MeV for the AME2020 testing set. It should be noted that most of the newly measured masses are far away from the valley of $\beta$-stability, indicating the reasonable predictive power in mass extrapolation. Besides, consistent behavior is obtained that the MTL-M12 model results in smaller rms deviation than the MTL-M9 ($0.397$ MeV). For charge radii, while only 616 charge radii are used in previous training process, there are about 25\% (153) newly measured data in the additional extrapolation set. However, the resultant rms deviations are just a little larger than those for the CR2013 testing set. The corresponding values are $\sigma_{\text{radius}}=0.026$~fm for MTL-M9 and $0.024$~fm for MTL-M12, respectively, indicating the good extrapolations of both models. Similarly, the MTL-M12 model still shows slightly better accuracy than the MTL-M9 model for the newly measured charge radii.

\begin{table}[htbp]
\caption{Rms deviations ($\sigma$) between model predictions and newly measured experimental data 
for nuclear masses and charge radii.}
\label{tab:newdata}
\centering
\begin{tabular}{lcc}
\toprule
Model   & $\sigma_{\text{mass}}$ (MeV) & $\sigma_{\text{radius}}$ (fm) \\
\midrule
MTL-M9  &  0.397  &  0.026  \\
MTL-M12 &  0.221  &  0.024  \\
\bottomrule
\end{tabular}
\end{table}
\subsubsection{Generalization with Garvey-Kelson relations}
\begin{figure}
 \includegraphics[width=7.5cm]{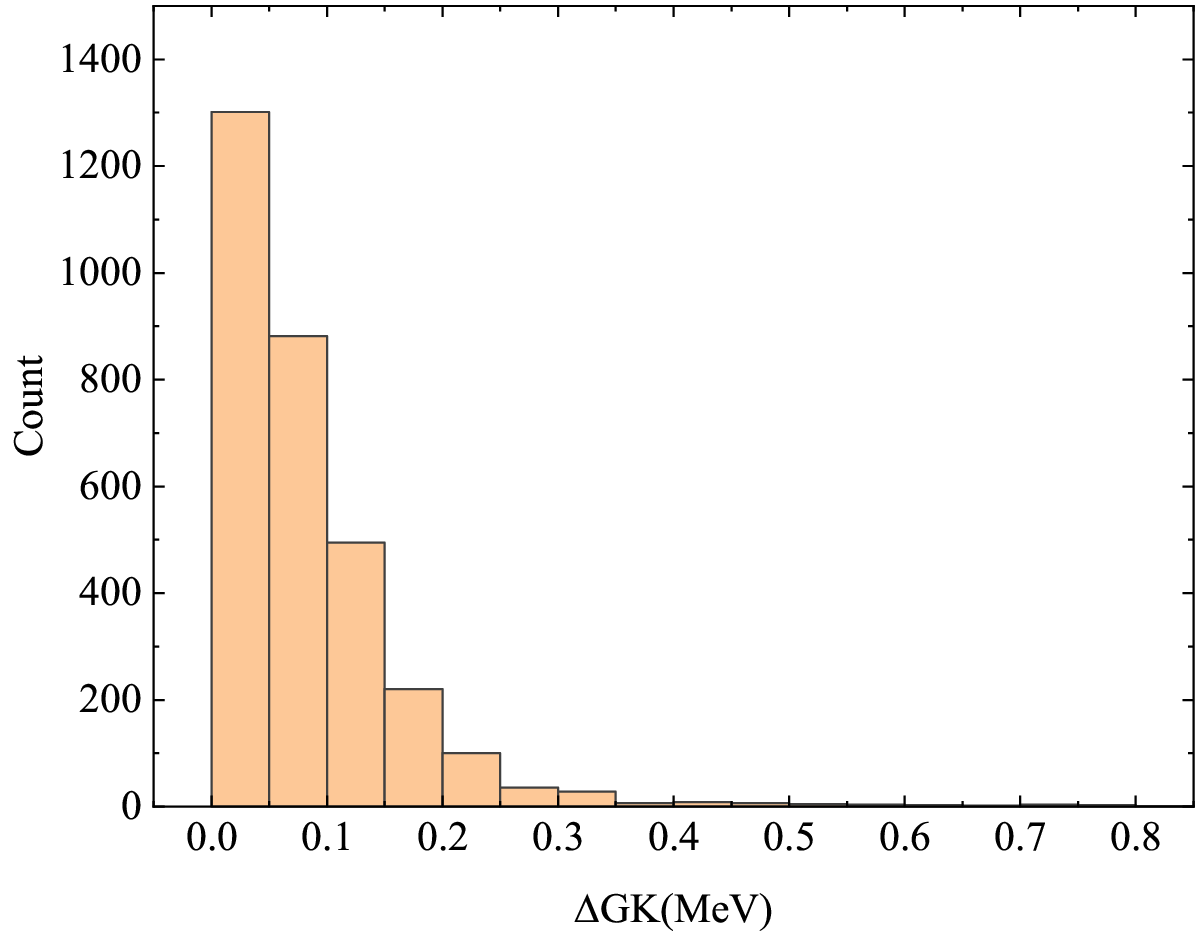}
 \caption{\label{figGK} Histogram of $\Delta GK =|GK^{\mathrm{ML}} -  GK^{\mathrm{exp}}\big|$ 
 for 3106 GK relations.}
\end{figure}

The GK relations combine the masses of neighboring nuclei through specific linear combinations, serving as a useful test of the consistency and reliability of nuclear mass models. Two commonly used GK relations \cite{Barea} are expressed as
\begin{equation}
\label{eq:GK}
GK(N,Z) =
\begin{cases}
\begin{aligned}
& M(N{+}2,Z{-}2) - M(N,Z) \\
& + M(N,Z{-}1) - M(N{+}1,Z{-}2) \\
& + M(N{+}1,Z) - M(N{+}2,Z{-}1),\\ 
& \text{for } N \ge Z, \\[8pt]
& M(N{-}2,Z{+}2) - M(N,Z) \\
& + M(N{-}1,Z) - M(N{-}2,Z{+}1) \\
& + M(N,Z{+}1) - M(N{-}1,Z{+}2), \\
& \text{for } N < Z ,
\end{aligned}
\end{cases}
\end{equation}
where $M(N,Z)$ is the mass of a nucleus with neutron number $N$ and proton number $Z$. In practice, we extracted a total of 3106 independent GK relations from Eq.~(\ref{eq:GK}) to serve as stringent tests. For each combination, we evaluate both the experimental value $GK^{\mathrm{exp}}$ and the ML-predicted value $GK^{\mathrm{ML}}$. The MTL-M12 dataset of nuclear masses is used to validate the model generalization, where the comparison is quantified by the difference $\Delta GK = \big| GK^{\mathrm{ML}} - GK^{\mathrm{exp}} \big|$. The value distribution for $\Delta GK$ is shown in Fig.~\ref{figGK}, where most values are within 0.2~MeV for more than 93\% of the independent GK relations. The average value for the quantity $\Delta GK$ is only 0.083 MeV, indicating the good consistency with the well-established GK relations and the good generalization of the present MTL-M12 model.

\subsection{Exploring task correlations with covariance matrices}
\label{correlation}
To further clarify the relationship between the predictions of nuclear mass and charge radius in the MTL model, we calculate their Pearson correlation coefficient from the learned task covariance matrix $B$ according to Eq.~(\ref{eq3}). For the M9 feature set, as listed in Table~\ref{tab1}, the positive Pearson correlation coefficient ($r_{\rm M9}=0.31$) demonstrates that the differences in nuclear mass and charge radius between experimental and theoretical values tend to increase or decrease together under joint training. For the M12 feature set, which is augmented by three shell-related terms, the correlation becomes even stronger ($r_{\rm M12}=0.57$). This indicates that the inclusion of shell-related features not only captures additional microscopic shell effects for both masses and radii, but also enhances the coupling between their predictions. The positive correlations for both M9 and M12 indicate that improvements in mass predictions tend to coincide with improvements in radius predictions, thereby reinforcing the advantage of joint learning. This is consistent with the mutual improvement in prediction accuracy achieved by the MTL model for both observables.

These results emphasize that correlations between the predictions of nuclear mass and charge radius depend sensitively on the selected feature sets. Macroscopic liquid-drop features already promote shared systematic trends and a positive correlation, while the additional shell-related features further strengthen this relationship. Hence, careful analysis of these correlations is meaningful for understanding and optimizing multi-task modeling.

\begin{table}[htbp]
\centering
\caption{Pearson correlation coefficients $r$ calculated by Eq.~(\ref{eq3}) for MTL models with 9 and 12 features.}
\label{tab1}
\begin{tabular}{lc}
\toprule
Model    & Pearson correlation coefficient $r$ \\
\midrule
MTL-M9   & 0.31 \\
MTL-M12  & 0.57 \\
\bottomrule
\end{tabular}
\end{table}

\subsection{Feature importance across different regions}
\label{importance}
To gain further insight into the physical mechanism underlying nuclear structure, we first divide the nuclear chart into three distinct regions: the neutron-rich region, the $\beta$-stable region, and the proton-rich region. Then the relative importances of different physical features for each region are systematically examined. We employ the SHapley Additive exPlanations (SHAP)~\cite{lundberg2017unified,ye2025understanding} method to quantify the differences, which can be used to interpret ML models by evaluating the feature importance for individual predictions. By calculating SHAP values for each nucleus and averaging them within each region, we can clearly identify the contribution tendency of specific physical features from the $\beta$-stable region toward either the neutron-rich or proton-rich extremes. The $\beta$-stable line is defined using the empirical relation:
\begin{equation}
Z_\beta(A) = \frac{A}{1.98 + 0.015\,A^{2/3}}.
\end{equation}
For each nucleus, we compute its distance from the stability line as
\begin{equation}
dZ = Z - Z_\beta(A).
\end{equation}
As adopted in Ref.~\cite{niu2019comparative}, a threshold of $dZ_{\mathrm{cut}} = 3$ is used to separate nuclei into three regions: $\beta$-stable region with $|dZ| \leq dZ_{\mathrm{cut}}$, proton-rich region with $ dZ > dZ_{\mathrm{cut}}$, and neutron-rich region with $ dZ < -dZ_{\mathrm{cut}}$. By separately analyzing the feature importance in each region, we can identify the feature relevance to nuclei under different isospin conditions more clearly.

\begin{figure}
 \includegraphics[width=8cm]{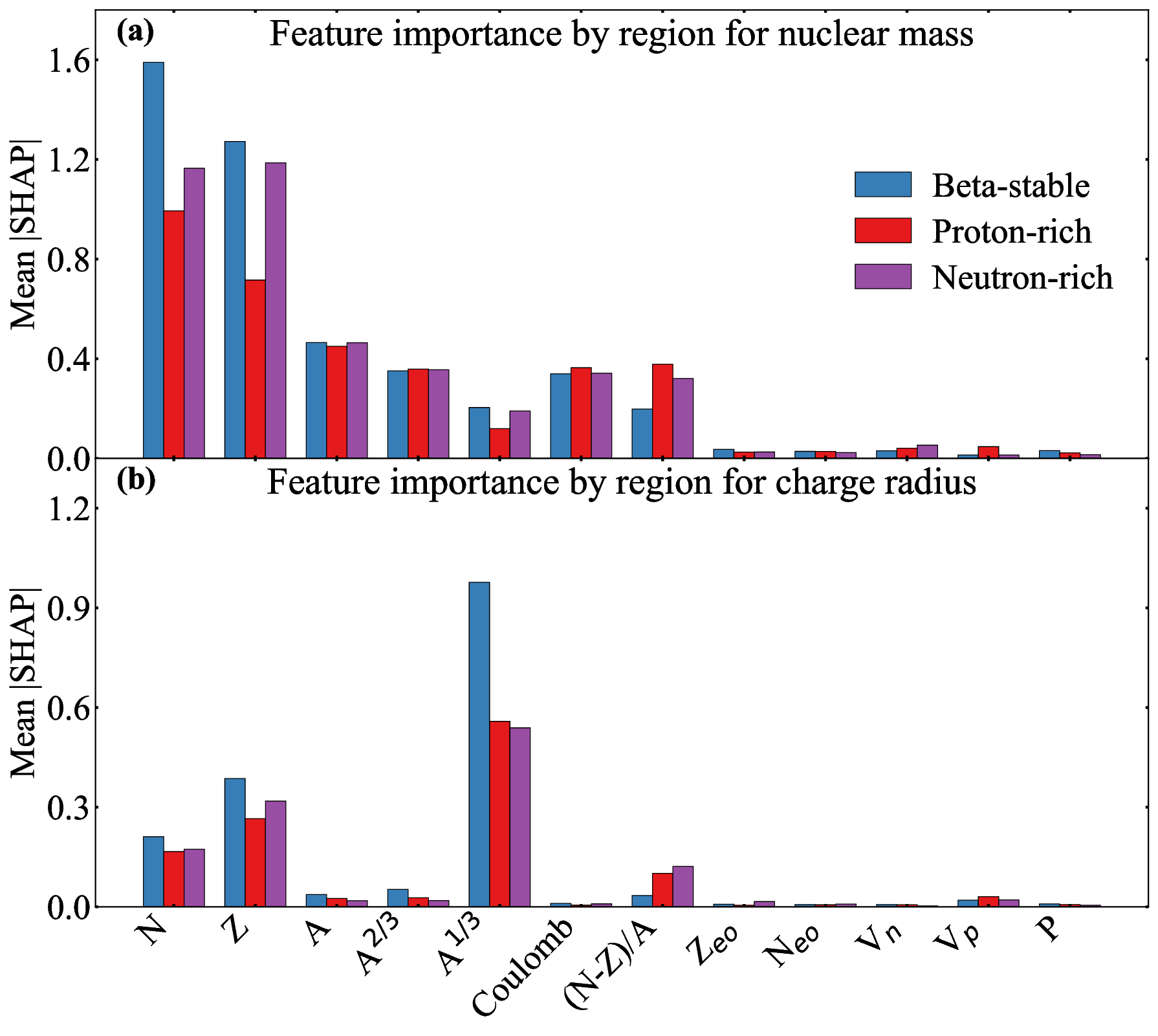}
\caption{\label{fig3} Feature importances for MTL-M12 in predicting nuclear mass shown in panel (a) and charge radius shown in panel (b) for $\beta$-stable, proton-rich, and neutron-rich regions, respectively. Each bar represents the mean absolute SHAP value, reflecting the importance of each physical feature in the specified nuclear region.}
\end{figure}

In Fig.~\ref{fig3}, we illustrate the importance of different physical features for predicting both nuclear masses and charge radii across the neutron-rich, $\beta$-stable, and proton-rich regions of the nuclear chart. As shown in panel (a) for the mass case, the bulk features namely neutron number $N$, proton number $Z$, volume term $A$, surface term $A^{2/3}$, Coulomb energy term, and isospin asymmetry term with generally larger SHAP values are demonstrated as the most important features, while the pairing term and shell-related terms show less importance, which is consistent with the well-established liquid-drop model. However, there are minor differences in three typical regions. For example, the isospin asymmetry term in both proton-rich and neutron-rich regions is much more important than in the $\beta$-stable region. This is because of the important additional symmetry energy for the nuclei in former two regions. Besides, for the proton-rich region, the SHAP values for the Coulomb term and proton shell effect ($v_{p}$) are larger than those for the $\beta$-stable region. This is due to the stronger electrostatic repulsion among protons and the enhanced role of proton-shell closures. In the neutron-rich region, the neutron shell effect ($v_{n}$) becomes more important compared to the $\beta$-stable region. Overall, these results demonstrate that bulk properties provide the primary contributions to the mass predictions in all regions of the nuclear chart, whereas the pairing and shell-related effects contribute less.

Unlike the feature importance for nuclear masses, the term $A^{1/3}$ corresponds to the largest SHAP value for the charge radius, along with the nucleon numbers $N$ and $Z$. Other features generally result in very small SHAP values, indicating minor importance for charge radius. This is due to the approximate saturation property of atomic nuclei, which is consistent with the well-known scaling relation of the charge radius proportional to $A^{1/3}$. In addition, the isospin asymmetry term $(N-Z)/A$ also exhibits important contributions with larger SHAP values in three typical regions. However, it is obvious that the $\beta$-stable region shows less important isospin asymmetric effects than the proton-rich and neutron-rich regions. These results indicate that the charge radii are primarily governed by the bulk $A^{1/3}$ dependence, with minor corrections from the isospin asymmetric effect and relatively weak contributions from other microscopic effects.

\section{Summary}
\label{summary}
In this work, a multi-task Gaussian process approach is introduced to jointly predict nuclear masses and charge radii using both 9 feature and 12 feature sets. The performances of multi-task learning (MTL) and single-task learning (STL) models for both nuclear mass and charge radius predictions are systematically evaluated. Using both 9 and 12 input features, the MTL models consistently lead to higher predictive accuracy for both nuclear masses and charge radii than STL models. Besides, consistent conclusion is also obtained that increasing the number of input features from 9 to 12 can also further improve both the STL and MTL performances. The MTL with 12 features achieves the smallest overall rms deviations of 0.136 MeV for nuclear mass and 0.007 fm for charge radius. These results suggest that the MTL not only provides a unified framework for jointly modeling nuclear mass and charge radius, but also improves the predictive accuracy for both observables across different feature sets.

Furthermore, three complementary tests are performed to examine the generalization and robustness of the MTL models. The first test, based on various train–test partitions, confirms the stability of the MTL predictions. The second test, involving extrapolations for newly reported nuclei far from stability, shows that the MTL retains reliable predictive accuracy beyond the training domain. The third test, based on the Garvey–Kelson relations, indicates that the MTL results are in good agreement with the local patterns among neighboring nuclei. These tests verify that the present MTL models possess good generalization capability.

In addition, covariance analysis reveals that the correlation between the predictions of nuclear mass and charge radius in the MTL is related to the choice of feature sets. Both the M9 and M12 feature sets exhibit positive correlations, indicating that the improvements in mass predictions tend to promote the improvements in radius predictions. 

Moreover, we apply interpretable SHapley Additive exPlanations (SHAP) techniques to analyze the feature importance for both nuclear mass and charge radius predictions across the $\beta$-stable, neutron-rich, and proton-rich regions. The SHAP results reveal that the bulk features, namely neutron number, proton number, volume term, surface term, Coulomb energy term, and isospin asymmetry term, are the primary factors for nuclear mass predictions across different regions. In particular, the isospin asymmetry term plays a more significant role in both proton-rich and neutron-rich regions compared to the $\beta$-stable region, while pairing effects generally contribute less. For charge radii, the $A^{1/3}$ dependence dominates, with moderate importance of the isospin asymmetry term $(N-Z)/A$ in the neutron-rich and proton-rich regions. These findings suggest that though bulk properties are the dominant factors for nuclear mass predictions, the charge radius is primarily governed by the $A^{1/3}$ dependence, with minor corrections from isospin asymmetry and shell effects for both predictions. These results demonstrate that the MTL models with physical features yield more accurate and interpretable predictions for nuclear observables.

\begin{acknowledgments}
This work was supported by the National Natural Science Foundation of China (Grants No. 12205105 and No. 12311540139), by the Fundamental Research Funds for the Central Universities (Grant No. 2024ZYGXZR058) and the startup funding of South China University of Technology.
\end{acknowledgments}

\bibliographystyle{apsrev4-2}
\bibliography{ref}
\end{document}